\documentclass[aps,pr,twocolumn,superscriptaddress,preprintnumbers,showpacs,amsmath,amssymb]{revtex4}

\usepackage{epsfig}
\usepackage{color}
\usepackage{graphicx}
\usepackage{dcolumn}
\usepackage{bm}

\newcommand{\BR}{{\cal B}}

\newcommand{\EE}{e^+e^-}

\newcommand{\pip}{\pi^+}
\newcommand{\pim}{\pi^-}
\newcommand{\piz}{\pi^0}

\newcommand{\ra}{\rightarrow}

\newcommand{\sumpt}{|\sum{\vec{p}_{t}^{\,*}}|}

\newcommand{\etac}{\eta_c(1S) }

\newcommand{\etato}{\eta \rightarrow }
\newcommand{\etap}{\eta^\prime }
\newcommand{\GG}{\gamma\gamma}
\newcommand{\etapto}{\eta^\prime \rightarrow }

\newcommand{\beq}{\begin{equation}}
\newcommand{\eeq}{\end{equation}}
\newcommand{\beqn}{\begin{eqnarray}}
\newcommand{\eeqn}{\end{eqnarray}}
\newcommand{\beqns}{\begin{eqnarray*}}
\newcommand{\eeqns}{\end{eqnarray*}}
\newcommand{\bfg}{\begin{figure}}
\newcommand{\efg}{\end{figure}}
\newcommand{\bitm}{\begin{itemize}}
\newcommand{\eitm}{\end{itemize}}
\newcommand{\bnum}{\begin{enumerate}}
\newcommand{\enum}{\end{enumerate}}
\newcommand{\btbl}{\begin{table}}
\newcommand{\etbl}{\end{table}}
\newcommand{\btbu}{\begin{tabular}}
\newcommand{\etbu}{\end{tabular}}

\begin{document}

\hspace{10mm}
\preprint{\vbox{ \hbox{   }
                \hbox{Belle Preprint 2010-17}
                \hbox{KEK Preprint 2010-26}
                \hbox{June 2012}
}}

\vspace{-10.0cm}
\title{\boldmath 
First study of $\etac$, $\eta(1760)$ and $X(1835)$ production via
$\etap\pip\pim$ final states in two-photon collisions 
}

\vspace{2.0cm}
\date{\today}
\vspace{0.5cm}

\begin{abstract}
The invariant mass spectrum of the $\etap\pip\pim$ final state produced
in two-photon collisions is obtained using a 673 fb$^{-1}$ data sample
collected in the vicinity of the $\Upsilon(4S)$ resonance with the 
Belle detector at the
KEKB asymmetric-energy $e^+e^-$ collider. 
We observe a clear signal of the $\etac$ and measure
its mass and width to be
$M(\etac)=(2982.7\pm 1.8(stat)\pm 2.2(syst)\pm 0.3(model))$ MeV$/c^2$ and 
$\Gamma(\etac) = (37.8^{+5.8}_{-5.3}(stat)\pm 2.8(syst) \pm 1.4(model))$ MeV$/c^2$.
The third error is an uncertainty
due to possible interference between the $\etac$ and a non-resonant component.
We also report the first evidence for $\eta(1760)$ decay to
$\etap\pip\pim$;
we find two solutions for its parameters, depending on the
inclusion or not of the $X(1835)$, whose existence is of marginal significance 
in our data.  
From a fit to the mass spectrum using coherent $X(1835)$ and $\eta(1760)$
resonant amplitudes, we 
set a $90\%$ confidence level upper limit on the product 
$\Gamma_{\gamma\gamma} \BR (\etap\pip\pim)$ for the $X(1835)$.
\end{abstract}

\affiliation{Budker Institute of Nuclear Physics SB RAS and Novosibirsk State University, Novosibirsk 630090}
\affiliation{Faculty of Mathematics and Physics, Charles University, Prague}
\affiliation{University of Cincinnati, Cincinnati, Ohio 45221}
\affiliation{Gifu University, Gifu}
\affiliation{Hanyang University, Seoul}
\affiliation{University of Hawaii, Honolulu, Hawaii 96822}
\affiliation{High Energy Accelerator Research Organization (KEK), Tsukuba}
\affiliation{Indian Institute of Technology Madras, Madras}
\affiliation{Institute of High Energy Physics, Chinese Academy of Sciences, Beijing}
\affiliation{Institute of High Energy Physics, Vienna}
\affiliation{Institute of High Energy Physics, Protvino}
\affiliation{Institute for Theoretical and Experimental Physics, Moscow}
\affiliation{J. Stefan Institute, Ljubljana}
\affiliation{Kanagawa University, Yokohama}
\affiliation{Institut f\"ur Experimentelle Kernphysik, Karlsruher Institut f\"ur Technologie, Karlsruhe}
\affiliation{Korea Institute of Science and Technology Information, Daejeon}
\affiliation{Korea University, Seoul}
\affiliation{Kyungpook National University, Taegu}
\affiliation{\'Ecole Polytechnique F\'ed\'erale de Lausanne (EPFL), Lausanne}
\affiliation{University of Maribor, Maribor}
\affiliation{Max-Planck-Institut f\"ur Physik, M\"unchen}
\affiliation{University of Melbourne, School of Physics, Victoria 3010}
\affiliation{Graduate School of Science, Nagoya University, Nagoya}
\affiliation{Kobayashi-Maskawa Institute, Nagoya University, Nagoya}
\affiliation{Nara Women's University, Nara}
\affiliation{National Central University, Chung-li}
\affiliation{Department of Physics, National Taiwan University, Taipei}
\affiliation{H. Niewodniczanski Institute of Nuclear Physics, Krakow}
\affiliation{Nippon Dental University, Niigata}
\affiliation{University of Nova Gorica, Nova Gorica}
\affiliation{Osaka City University, Osaka}
\affiliation{Pacific Northwest National Laboratory, Richland, Washington 99352}
\affiliation{Peking University, Beijing}
\affiliation{University of Science and Technology of China, Hefei}
\affiliation{Seoul National University, Seoul}
\affiliation{Sungkyunkwan University, Suwon}
\affiliation{School of Physics, University of Sydney, NSW 2006}
\affiliation{Tata Institute of Fundamental Research, Mumbai}
\affiliation{Excellence Cluster Universe, Technische Universit\"at M\"unchen, Garching}
\affiliation{Tohoku University, Sendai}
\affiliation{Department of Physics, University of Tokyo, Tokyo}
\affiliation{Tokyo Institute of Technology, Tokyo}
\affiliation{Tokyo Metropolitan University, Tokyo}
\affiliation{CNP, Virginia Polytechnic Institute and State University, Blacksburg, Virginia 24061}
\affiliation{Yonsei University, Seoul}
  \author{C.~C.~Zhang}\affiliation{Institute of High Energy Physics, Chinese Academy of Sciences, Beijing} 
  \author{H.~Aihara}\affiliation{Department of Physics, University of Tokyo, Tokyo} 
  \author{D.~M.~Asner}\affiliation{Pacific Northwest National Laboratory, Richland, Washington 99352} 
  \author{T.~Aushev}\affiliation{Institute for Theoretical and Experimental Physics, Moscow} 
  \author{A.~M.~Bakich}\affiliation{School of Physics, University of Sydney, NSW 2006} 
  \author{Y.~Ban}\affiliation{Peking University, Beijing} 
  \author{K.~Belous}\affiliation{Institute of High Energy Physics, Protvino} 
  \author{M.~Bischofberger}\affiliation{Nara Women's University, Nara} 
  \author{T.~E.~Browder}\affiliation{University of Hawaii, Honolulu, Hawaii 96822} 
  \author{A.~Chen}\affiliation{National Central University, Chung-li} 
% ...
\collaboration{The Belle Collaboration}

\maketitle

\section{Introduction}
As the lowest charmonium state, the $\etac$ meson plays an important role 
in tests of QCD. However, even its main parameters, such as the mass, width and
two-photon width, have not been well measured and the measurements that have 
been reported show a large scatter 
of values~\cite{pdg10}. 
Discrepancies among measurements for the $\etac$ product of the two-photon
width and decay branching fraction into four-meson final states were confirmed 
earlier~\cite{Charmonium two-photon}.
A recent measurement of the $\etac$ that found a significant 
interference between the $\etac$ and the non-resonant 
background~\cite{BES etac} may have
clarified the reason for discrepancies among
$\etac$ parameter measurements~\cite{bram}. 
Significant model-dependent uncertainty in the measurement of the $\etac$ product 
branching fractions due to interference between the $\etac$ and a non-resonant 
component has also been studied in $B\ra K \etac$ decays~\cite{vinokur}.

The $X(1835)$ resonance was observed and confirmed recently by the 
BES collaboration 
in $J/\psi\ra\gamma X(1835)$ decays 
where $X(1835)\ra\etap\pip\pim$~\cite{BES X1835}, with
mass $M=(1836.5\pm 3.0^{+5.6}_{-2.1})$ MeV$/c^2$ and
width $\Gamma=(190\pm 9^{+38}_{-36})$ MeV$/c^2$.
A variety of speculations on the nature of the $X(1835)$ have been reported,
including baryonium~\cite{S.L. Shi and C.S. Gao} with 
sizable gluon content~\cite{G.J Ding and M.L. Yan},
glueball~\cite{N. Kochelev and D. Min, Bin An Li, X.G. He}, and
a radial excitation of the $\eta^\prime$~\cite{Tao Huang, E Klempt}. 
The BES experiment has suggested that the $X(1835)$ may be related to the
$p \overline{p}$ threshold enhancement seen
in $J/\psi\ra\gamma p\overline{p}$ 
decays~\cite{BES ppbar enhancement, BESIII x1835 ppbar}.
An additional structure, the $\eta(1760)$, was observed in the radiative 
$J/\psi$ decays to 
$\gamma\rho\rho$ and $\gamma\omega\omega$ by MARKIII~\cite{MK3 eta1760} and 
DM2~\cite{DM2 eta1760} and to $\gamma\omega\omega$ and $\gamma\eta\pi^+\pi^-$ 
by BES~\cite{BES eta1760}.
The $\eta(1760)$ state has been proposed as a mixture
of a gluonic meson with a conventional $q\bar{q}$ state~\cite{Page and Li},
rather than a pure $q\overline{q}$ meson, and this hypothesis is supported 
by a BES analysis of 
$J/\psi\ra\gamma\omega\omega$ decays~\cite{BES eta1760}. 
Hence, an investigation of the nature of both the $X(1835)$ 
and $\eta(1760)$ is of 
interest~\cite{J.L. Rosner}.
In radiative $J/\psi$ decays, hadrons are produced via two gluons; thus,
the production of final states with a gluon-enriched component is expected
to be enhanced. 
In light of the similar structure of the two-photon and two-gluon couplings, 
a comparison of the $\GG$ width of a meson to its production rate in
radiative $J/\psi$ decays can provide information on its quark 
and gluon composition. 
The two-photon coupling to the gluonic component is expected to be 
very weak so
measurements of two-photon widths can help clarify the nature of the $X(1835)$ 
and $\eta(1760)$.

In this paper, we report the first observation of $\etap\pip\pim$
production in two-photon collisions using a 673 fb$^{-1}$ data sample
(605 fb$^{-1}$ on the 
$\Upsilon(4S)$ resonance and 68 fb$^{-1}$ at 
60 MeV below the resonance) accumulated with the Belle 
detector~\cite{Belle detector} at the KEKB asymmetric-energy $\EE$
collider~\cite{KEKB collider}. We measure parameters of the $\etac$,
provide first evidence for $\eta(1760)\ra\etap\pip\pim$ decay,
and give limits on the two-photon production of the $X(1835)$.

\section{Detector and Monte Carlo simulation}\label{Detector}
The Belle detector
is a large-solid-angle magnetic spectrometer that consists of a
silicon vertex detector, a 50-layer central drift chamber (CDC),
an array of aerogel threshold Cherenkov counters (ACC), a barrel-like
arrangement of time-of-flight scintillation counters (TOF), and an
electromagnetic calorimeter comprised of CsI (Tl) crystals (ECL).
These detectors are located inside a superconducting solenoid coil
that provides a 1.5 T magnetic field. An iron flux return located outside 
the coil is instrumented to detect $K_L^0$ mesons and to identify 
muons~\cite{Belle detector}.

Monte Carlo (MC) events of the two-photon process $\gamma^*\gamma^* 
\rightarrow \etap\pip\pim$ are generated with the TREPS 
code~\cite{TREPS MC code} based on an Equivalent Photon Approximation
(EPA)~\cite{Lgg BW for 2gam},
where the $\etap$ decays generically according to the JETSET7.3 decay table
~\cite{JETSET}.  
An isotropic phase space distribution is assumed for $\etac$, $\eta(1760)$ 
and $X(1835)$ decays to the three-body $\etap\pip\pim$ final state.
The GEANT-based simulation package~\cite{Belle MC simulation}
with trigger conditions included is employed for the propagation of the 
generated particles through the Belle detector.

\section{Event selection}\label{event selection}
The $\etac$, $\eta(1760)$ and $X(1835)$ (collectively denoted as $R$) 
candidates are reconstructed from the decay chain 
$R \ra \etap\pip\pim$, $\etapto\eta\pip\pim$,
and $\etato \gamma\gamma$. Two photons and two 
$\pip\pim$ pairs are detected in the final state.

\subsection{Selection criteria}
At least two neutral clusters and four charged tracks with zero net charge 
are required in each event. Candidate photons are neutral clusters that 
have an energy deposit greater than 100 MeV in the ECL 
and are not near any of the charged tracks.
The polar angle of the charged tracks, ${\it i.e.}$, the angle with respect to 
the direction opposite the positron beam axis in the laboratory system,
must satisfy $\cos\theta\in[-0.8660,+0.9563]$. 
To enhance the detection efficiency for low momentum charged tracks, loose 
requirements on the impact parameters perpendicular to ($dr$) and along ($dz$)
the beam line from the interaction point are applied: 
$dr < 5$ ($<3$,~$<2$,~$<1$) cm and $|dz| < 5$ ($<5$,~$<4$,~$<3$) cm 
for the track transverse
momentum $p_t<0.2$ ($\in[0.2,0.3]$,~$\in[0.3,0.4]$,~$>0.4$) GeV$/c$.
The scalar sum of the absolute momenta for all the charged tracks
and neutral clusters   
and the sum of the ECL cluster energies in the laboratory system 
are required to be $p_{\rm sum} < 5.0$~($<5.5$) GeV$/c$ 
for the $\etap\pip\pim$ system in the mass region below $2.7$ GeV$/c^2$
(in the $\etac$ region) and $E_{\rm sum} < 4.5$ GeV.

Events with an identified kaon ($K^\pm$ or $K_S^0\ra\pip\pim$) or proton 
are vetoed. 
For charged tracks, information from the ACC, TOF
and CDC is combined to form a likelihood $\mathcal{L}$ for 
hadron identification. 
A charged track with the likelihood ratio of
$\mathcal{L}_K/(\mathcal{L}_\pi + \mathcal{L}_K)>0.8$ 
is identified as a kaon; one
with $\mathcal{L}_\pi/(\mathcal{L}_\pi + \mathcal{L}_K)>0.2$ as a pion. 
With these loose requirements, the efficiency for pion identification 
is about $99\%$. 
A proton is identified by the requirement 
$\mathcal{L}_p/(\mathcal{L}_p + \mathcal{L}_K)>0.95$. 
$K_S^0$ candidates
are reconstructed from a pair of charged pion tracks with invariant
mass within 16 MeV$/c^2$ ($3\sigma$) of the nominal $K_S^0$ mass.

The $\eta$ from $\etap\ra\eta\pip\pim$ decay is reconstructed 
via its two-photon decay mode, where the two-photon invariant mass 
is in the window
$M_{\gamma\gamma}\in[0.524,0.572]$ GeV$/c^2$ 
($\pm 2\sigma$ of the nominal $\eta$ mass). 
To suppress background photons from $\piz$ decay, we exclude any photon that,
in combination with another photon in the event, has an invariant mass within 
the window $|M_{\gamma\gamma}-m_{\piz}|<18$ MeV$/c^2$. 
The two-photon-energy asymmetry,
$A_{\rm sym} = |E_{\gamma 1}-E_{\gamma 2}|/(E_{\gamma 1}+E_{\gamma 2})$,
is required to be less than 0.8 to suppress the fake $\eta$
combinatorial background.
The $\etap$ candidate is reconstructed from
the $\eta$ candidate and the $\pip\pim$ track pair that results in 
an invariant mass within 
$M_{\eta\pip\pim}\in[0.951,0.963]$ GeV$/c^2$ 
($\pm 2\sigma$ of the nominal $\etap$ mass).
To improve the momentum resolution of the $\eta$ and $\etap$, 
a mass-constrained fit to the $\eta$ and
two separate fits to the $\etap$ (one with a constrained vertex and the other with 
the mass constrained to the $\etap$)
are applied. 

The $\etap\pip\pim$ candidates are reconstructed by combining the
$\etap$ candidate and the remaining $\pip\pim$ track pair. 
For multi-candidate events, the candidate with the smallest $\chi^2_m$
from the $\etap$ mass-constrained fit is selected. For $\etap\pip\pim$
combinations with 
invariant mass $W = 1.84~(2.98)$ GeV$/c^2$, $19\%~(7.1\%)$ of the signal MC events
have more than one candidate
per event, from which the correct candidate is selected 
$98\%~(91\%)$ of the time.

\subsection{Background and optimization for $\sumpt$ requirement}
Signal and non-resonant events can be 
produced in two-photon collisions
via the processes $\EE \to \EE R$ and $\EE \to \EE \etap\pip\pim$, respectively, 
where quasi-real photons are emitted from the beam $e^+$ and $e^-$ particles
at small angles with respect to the beam line. 
These events
tend to carry small transverse momentum $\sumpt$,
which is determined by taking the absolute value of the vector sum of the 
transverse momenta of $\etap$ and the $\pip\pim$ tracks 
in the $e^+e^-$ center-of-mass system.

The $\eta'$-sideband, denoted $\eta'$-$sdb$, arises from $\eta\pip\pim\pip\pim$ 
and $\gamma\gamma\pip\pim\pip\pim$ (without $\eta$) 
combinations that survive the $\etap$ selection criteria 
except that the $\eta\pip\pim$ combination whose mass is nearest that of the $\etap$
lies between 0.914 and 0.934 GeV/$c^2$ or between 0.98 and 1.0 GeV/$c^2$.  
Similar events with an $\eta\pi\pi$ mass within the $\etap$ acceptance window form 
a featureless background denoted $b_1$ in the $R$-candidate sample. 
The $\eta'\pi^+\pi^-X$ background, denoted $b_2$, has additional particles 
in the event beyond the $R$ candidate.
Other non-exclusive backgrounds, including those arising 
from initial state radiation, are found to be negligible.

Significant background reduction is achieved 
by applying a $\sumpt$  requirement.
The $\sumpt$ distribution for the signal peaks 
at small values, while that for both backgrounds
decreases toward $|\sum{\vec{p}_{t}^{\,*}}| = 0$
due to vanishing phase space~\cite{backg any}.

The $\etac$ state is well 
established~\cite{BES X1835, Charmonium two-photon, etac and etappipi} 
and its signal yield in our data sample is large. 
We utilize a control sample of $\etap\pip\pim$ candidates from half the data, 
with $W$ between 2.6 and 3.4 GeV$/c^2$,
to establish the $\sumpt$ requirement
under the assumption that the $\sumpt$ distribution 
is similar for events with $W < 2.2~$ GeV$/c^2$.
The $\etap$-$sdb$ events from the full data 
sample are added to this control sample 
under the assumption that their $\sumpt$ distribution is similar 
to that of the $b_1$ 
background so that the signal fraction 
in this control sample is close to that 
in the $W$ mass region below $2.2$ GeV$/c^2$
in the full data sample.
We use the relative statistical 
error for the $\etac$ yield in fitting the $\etap\pip\pim$ mass spectra 
to optimize the $\sumpt$ requirement.
The requirement $|\sum{\vec{p}_{t}^{\,*}}| < 0.09~$ GeV$/c$ 
($p_t$-balanced) is applied to the $R$-candidate sample
since it minimizes this relative error.

\section{Background estimation}\label{sec Background}
\begin{figure}[htbp]
\centerline{
\hbox{\psfig{file=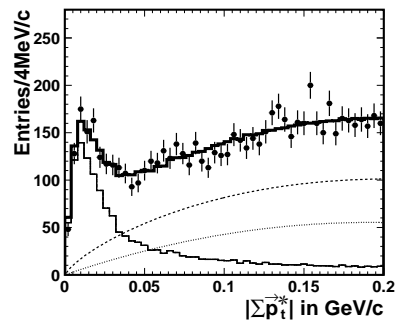,height=4.3cm,width=5.3cm}}}
\caption{The $\sumpt$ distributions for 
the mass region below $2.2$ GeV$/c^2$ 
the data sample. 
The data points with error bars are from the $\eta'\pip\pim$-candidate sample 
before the $|\sum\vec{p}_t^*|$ requirement, the thick-solid histogram is the best fit, 
the thin solid histogram is the signal component, the thin-dashed curve is 
the $b_1$ component (whose shape is taken from the $\etap$-$sdb$ sample), 
and the thin-dotted curve is the $b_2$ component.
} \label{ptcms distribution}
\end{figure}

The $b_1$ component in the $\etap\pip\pim$ mass and $\sumpt$ distributions
are determined in the fits to the $\etap$-$sdb$ events (normalized)
in the $p_t$-balanced and $p_t$-unbalanced (see below) samples, respectively. 
The residual $b_2$ component in the final $R$-candidate sample
can be separated using the $\sumpt$ distribution. By doing so, its distribution in
$\etap\pip\pim$ mass is determined.
Figure~\ref{ptcms distribution} shows the $\sumpt$ distribution
for signal MC events and data in the mass region below $2.2$ GeV$/c^2$.

A $p_t$-unbalanced data subsample, in which
the backgrounds dominate over the signal, is selected with the requirement
$\sumpt \in[0.15,0.2]$ GeV$/c$.
The $\eta' \pi \pi$ mass distribution of this $p_t$-unbalanced subsample 
is fit to two separate background functions, one for the $b_1$ component
with its yield and shape fixed at the values determined using the corresponding 
$\etap$-$sdb$ sample
and the other for the $b_2$ component with its yield $y_{unbal}$ 
and shape parameters allowed to float.
We use the same shape for the $b_2$ component in the later fit to the 
$\eta' \pip \pim$ mass spectrum for the final $R$-candidate sample.
Here, the assumption of the same shape in the invariant mass distribution for 
the $b_2$ component in the $p_t$-balanced and -unbalanced samples is implied. 
In the fit shown in Fig.~\ref{ptcms distribution}, 
the signal function for $R$ and non-resonant events
is defined by a histogram of the signal MC events 
with its shape parameters fixed but yield floated; 
the $b_1$ component is described by a threshold function with its yield and 
shape parameters fixed; 
the $b_2$ component is described by a quadratic function 
with its yield and shape parameters floated.
Here, the quadratic function for the $b_2$ is constrained to 
the origin, since $b_2$ background events selected as $\etap \pip\pim$ 
with missing $X$ should have non-zero transverse momentum.
From the fit, we obtain the $b_2$ yields 
in the $p_t$-balanced and -unbalanced subsamples; the ratio of 
these yields is $y_{bal}/y_{unbal}~=~0.723 \pm 0.043$.
(The corresponding $b_2$ yield 
ratio for $2.6\,{\rm GeV}/c^2 < W < 3.4\,{\rm GeV}/c^2$ is $0.93 \pm 0.11$.)
The $b_2$ yield $y^\prime_{bal}$ in the $\etap \pip\pim$ mass spectrum 
for the final $R$-candidate sample is obtained from the yield $y^\prime_{unbal}$
scaled to this ratio.

\begin{figure}[htbp]
\centerline{\hbox{\psfig{file=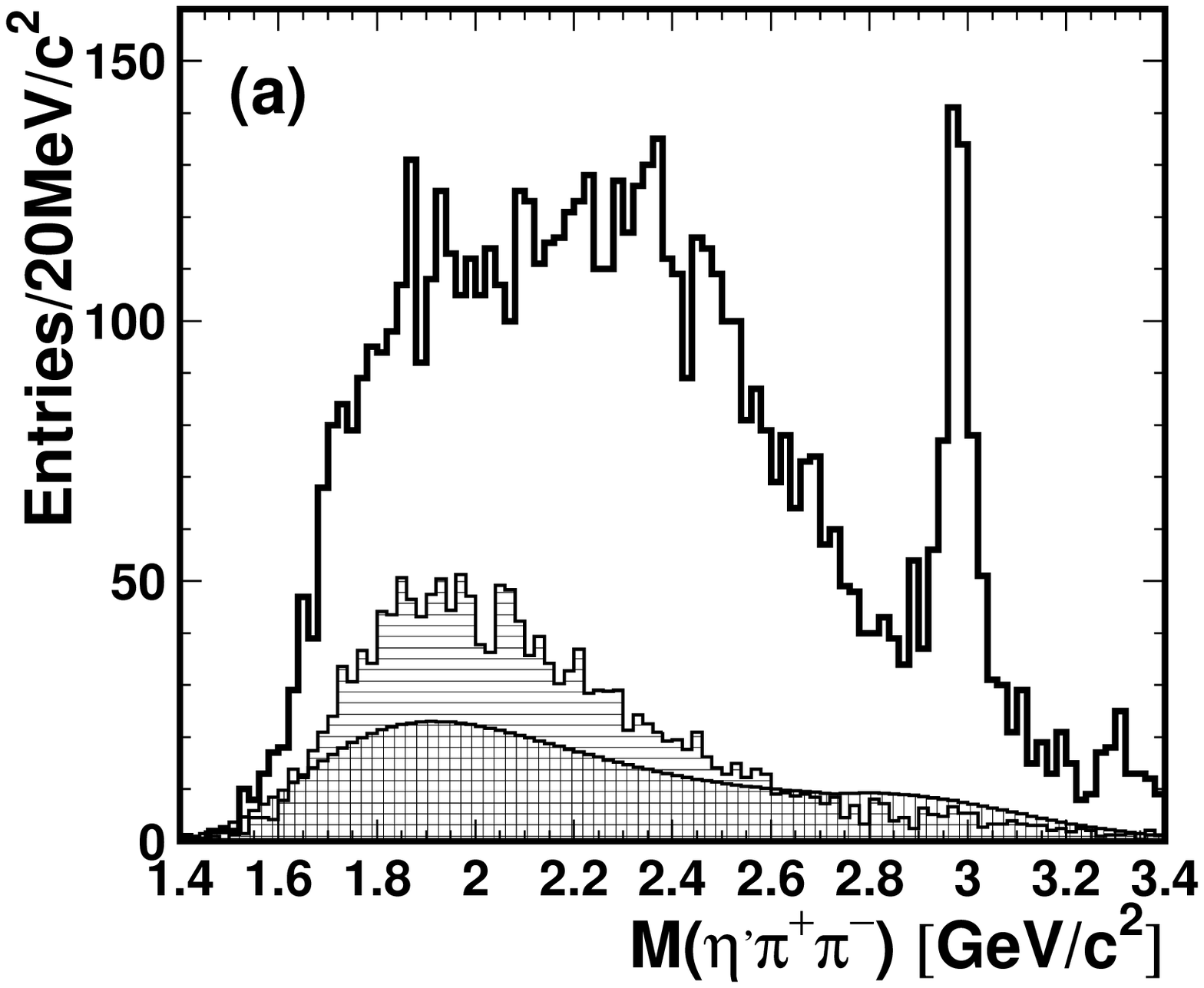,height=4.3cm,width=5.3cm}}}
\centerline{\hbox{\psfig{file=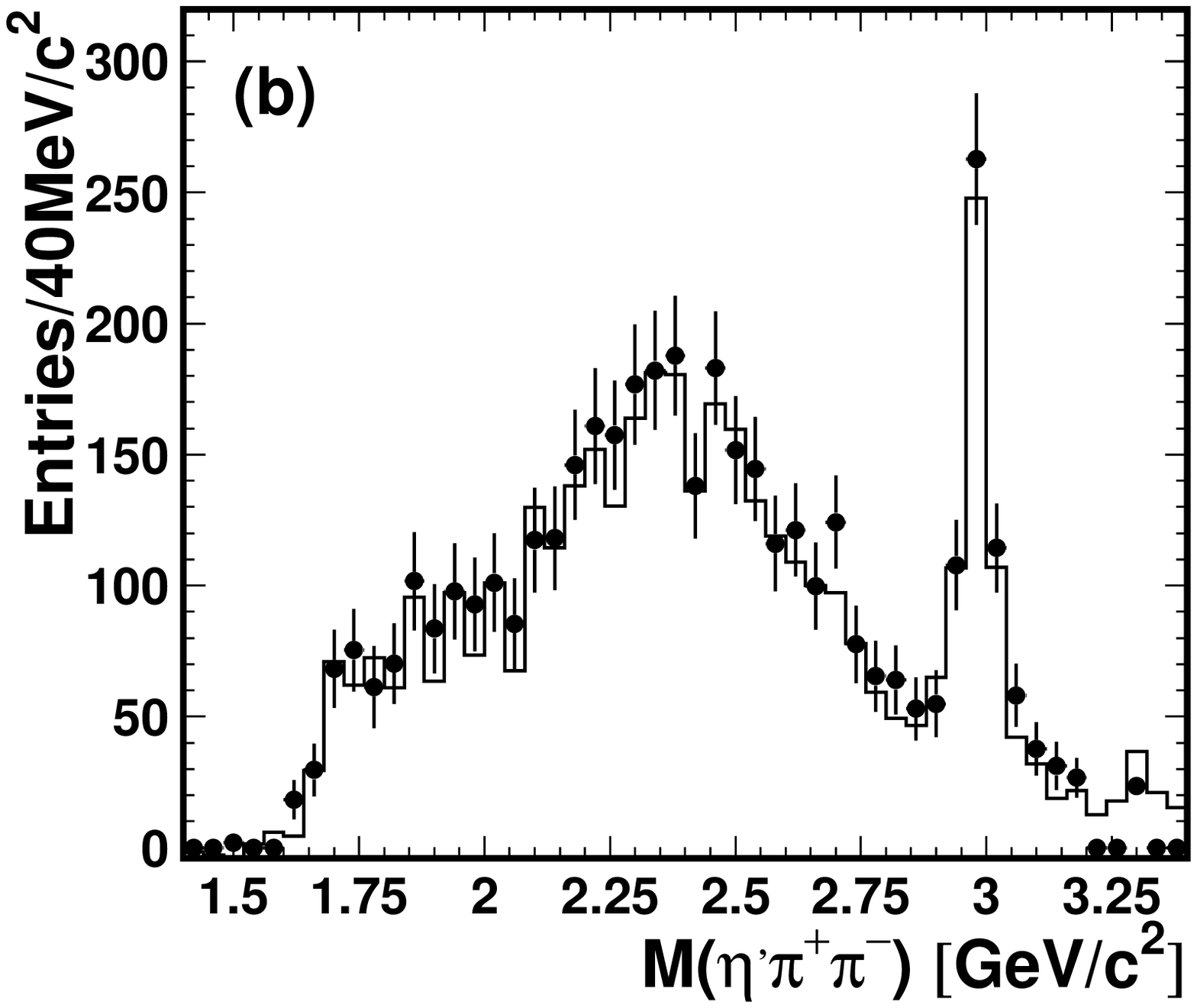,height=4.3cm,width=5.3cm}}}
\caption{Invariant mass distribution for the $\etap\pip\pim$ 
candidates. (a) The open histogram represents the data;
the horizontal (vertically) hatched histogram is the contribution from
the $b_1$ ($b_2$) background.
(b) The histogram shows the data after subtraction of both the $b_1$ 
and $b_2$ background components; the points with error bars
are the $\etap\pip\pim$ yields extracted from 
fitting the $\sumpt$ distribution in each sliced mass bin.
} \label{mass for etap pipi}
\end{figure}

The invariant mass distributions for the $\etap\pip\pim$ candidates, 
as well as those for
the $b_1$ and $b_2$ backgrounds, 
are shown in Fig.~\ref{mass for etap pipi}. 
In addition to the prominent $\etac$ signal, 
an enhanced shoulder 
is evident in the mass region below $2$ GeV$/c^2$
in the $b_1$- and $b_2$-subtracted histogram of Fig.~\ref{mass for etap pipi}(b).
The robust enhancement is also seen in the 
$\etap\pip\pim$ yields
extracted from fitting the $\sumpt$ distributions 
in each sliced mass bin, shown as data points 
with error bars in Fig.~\ref{mass for etap pipi}(b).

\section{Fitting mass spectrum}
The cross section of $R$ production in the two-photon 
process $\EE \to \EE R$ is approximated by

\begin{eqnarray}
\sigma(\EE \ra \EE R) = \int \sigma_{\GG \ra R}(W) 
\frac{dL_{\gamma\gamma}} {dW} dW,  
\label{cross section 1}
\end{eqnarray}
where the two-photon luminosity function $\frac{dL_{\GG}}{dW}$  
is calculated in the EPA using TREPS and the cross section 
$\sigma_{\GG\ra R}(W)$ for $C$-even resonance production 
with zero spin is described by a Breit-Wigner ($BW$) function
$f_{BW}(W)$~\cite{Lgg BW for 2gam}:

\begin{eqnarray}
\sigma_{\gamma\gamma\ra R}(W) &=& f_{BW}(W)\cdot\Gamma_{\GG}   \nonumber \\
&=& \frac{8\pi\Gamma\cdot\Gamma_{\GG}} {(W^2- M^2)^2 + \Gamma^2 M^2}, 
\label{cross section 2}
\end{eqnarray}
where $M$, $\Gamma$ and $\Gamma_{\gamma\gamma}$ are the mass, total width and
two-photon decay width of the $R$, respectively.

The signal yield $n_s$, $M$ and $\Gamma$ are extracted by maximizing the extended 
likelihood function,
\begin{eqnarray}
\mathcal{L} &=& \frac{e^{-(n_s + \sum_{k=1}^{3} {n_{b,k}} )}}{N!} \prod_{i=1}^{N} [
\mathit{n_s\cdot f_s(u_i; M,\Gamma)} \nonumber \\
&& + \sum_{k=1}^{3} \mathit{n_{b,k} \cdot f_{b,k}(u_i; p_{b,k})} ],
\label{Likelihood func}
\end{eqnarray}
where $n_s$ ($n_{b,k}$) is the number of signal ($k$-th background component) events, 
$N$ is the total number of candidate events, $i$ is the event identifier 
and $u_i$ is the measured invariant mass
for the $i$-th event. 
The probability density function (PDF) $f_s$ for the $R$ signal 
is a $BW$ function convolved with mass resolution
after corrections for $\frac{dL_{\GG}}{dW}$ and the efficiency.
The $k$-th background's PDF and its parameters are denoted by
$f_{b,k}$ and $p_{b,k}$, respectively.
In the fit, $n_s$, $M$ and $\Gamma$ for the signal are allowed to float
unless stated otherwise;
$n_{b,k}$ and $p_{b,k}$ for non-resonant background ($NR$) are allowed to 
float while those for the $b_1$ and $b_2$ backgrounds are fixed.
Two distinct fits are performed: in the lower mass region
$1.4\,{\rm GeV}/c^2 < W < 2.7\,{\rm GeV}/c^2$ where the $NR$ 
(as well as $b_1$ and $b_2$) background component is described by 
a threshold function~\cite{mnfit} with a reasonable description of the threshold effect,
and in the higher mass region
$2.6\,{\rm GeV}/c^2 < W < 3.4\,{\rm GeV}/c^2$ (near the $\eta_c(1S)$)
where all the background components are described by 
an exponential of a third-order polynomial. 

The evaluation of the significance of any marginal $R$ signal
in the lower-mass fit is sensitive to the assumed 
background shape.
We have examined results of various fits with different descriptions 
of the background:
(1) one threshold function for a sum of all three background components
({\it i.e.}, $b_1$, $b_2$ and $NR$); (2) two separate threshold 
functions, one for $b_1$
and the other for $b_2$ plus $NR$; (3) three separate threshold 
functions, one each
for $b_1$, $b_2$ and $NR$, respectively; (4-6) the three 
background functions defined above, in each case convolved with a mass resolution function 
after corrections for the two-photon luminosity and efficiency. 
We fit the $\etap\pip\pim$ mass spectrum for a possible 
$\eta(1760)$ signal in the mass region below $2.7$~GeV$/c^2$ using the six
different background models described above.
Option (3) provides the smallest statistical significance for a signal resonance, 
and is conservatively chosen for the background description. 

The product of the two-photon decay width and the $\eta^\prime \pip\pim$ branching fraction 
for the $R$ is determined as:
\begin{eqnarray}
\Gamma_{\gamma\gamma} \BR (R\ra \eta^\prime \pip\pim) = ~~~~~~~~~~~~~~~~~~~~~~~~~ \nonumber \\
 \frac {n_{s}} {L_{int}\cdot \int f_{BW}(W) \frac{dL_{\gamma\gamma}(W)}{dW} \epsilon(W)dW },
\label{Eq1 prd of Br and 2g-width}
\end{eqnarray}
where the efficiency $\epsilon$ includes the branching fractions for 
$\mathcal{B}(\eta^\prime \ra \eta\pip\pim)$ 
and $\mathcal{B}(\eta\ra\gamma\gamma)$.

\subsection{Results of the $\eta(1760)$ fit}
\begin{figure}[htbp]
\centerline{
\hbox{\psfig{file=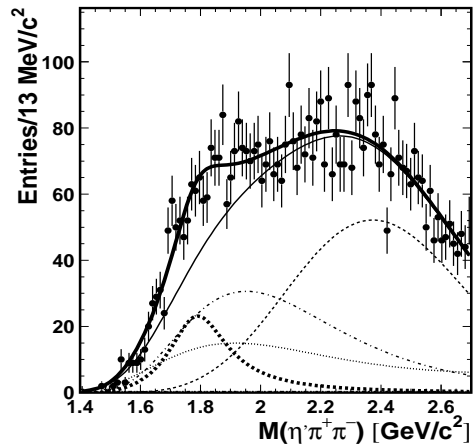,height=6.0cm,width=6.5cm}}
}
\caption{
The invariant mass distribution for $\etap\pip\pim$ candidates in
the lower-mass region.
The points with error bars are data. The thick solid line is the best fit;
the thin solid line is the 
total background. The thick dashed line is the fitted signal for the 
$\eta(1760)$. The thin dashed, dot-dashed and dotted lines are the $NR$,
$b_1$ and $b_2$ background components, respectively.}\label{Fig fit to eta1760}
\end{figure}

We assume that only one resonance is produced in the mass range below 2.7~GeV/$c^2$ 
and that there is no interference between the signal and $NR$ components. 
Figure~\ref{Fig fit to eta1760} shows the results of the fit
for the decay $R \ra\etap\pip\pim$.
A signal with a yield $n_s = 465^{+131}_{-124}$ and
a statistical significance of $4.8\sigma$ is found with
mass $ M=(1768^{+24}_{-25})$ MeV$/c^2$ and 
width $\Gamma = (224^{+62}_{-56})$ MeV$/c^2$; we denote this as $\eta(1760)$.
The statistical significance, in units of standard deviation ($\sigma$), is calculated 
using the $\chi^2$ distribution $-2\cdot {\rm ln}(\mathcal{L}_0/\mathcal{L}_{max})$
with $N_{\rm dof}$ degrees of freedom. Here, 
$\mathcal{L}_{max}$ and $\mathcal{L}_0$ denote the maximum likelihood 
with the signal yield floating and fixed at zero, respectively, and
$N_{\rm dof}=3$ is the difference in the number of floating parameters between
the nominal fit and the fit with the signal yield fixed at zero.
The product of the two-photon decay width and branching fraction is 
determined to be 
$\Gamma_{\GG} \mathcal{B}(\eta(1760)\ra\etap\pip\pim) = 
(28.2^{+7.9}_{-7.5})$ eV$/c^2$.

\subsection{Results of the $X(1835)$ fit}\label{Fit result for X(1835)}
According to existing observations~\cite{BES X1835, pdg10}, two resonances, 
$X(1835)$ and $\eta(1760)$, have been reported
in the lower mass region above the $\eta^\prime\pip\pim$ threshold.
Assuming that both $X(1835)$ and  $\eta(1760)$ have the same spin-parity 
of $J^{PC}=0^{-+}$, the effect of interference
between these two states must be taken into account in 
any attempt to extract a signal yield for the $X(1835)$. 
Each resonance is described by a $BW$ amplitude:

\begin{eqnarray}
{g_{BW}(W) = \frac{1}{(W^2-M^2) + i\Gamma {M}}},
\label{BW function}
\end{eqnarray}
and the amplitude for the two interfering resonances 
is written as
\begin{eqnarray}
\mathcal{M}(W) = A_1\cdot g_{BW1}(W) + A_2\cdot g_{BW2}(W) \cdot e^{i\phi},
\label{BW amplitude 2}
\end{eqnarray}
where $\phi$ is the relative phase between the two resonances 
and $A_1$ and $A_2$ are normalization factors.

\begin{figure}[htbp]
\centerline{\hbox{\psfig{file=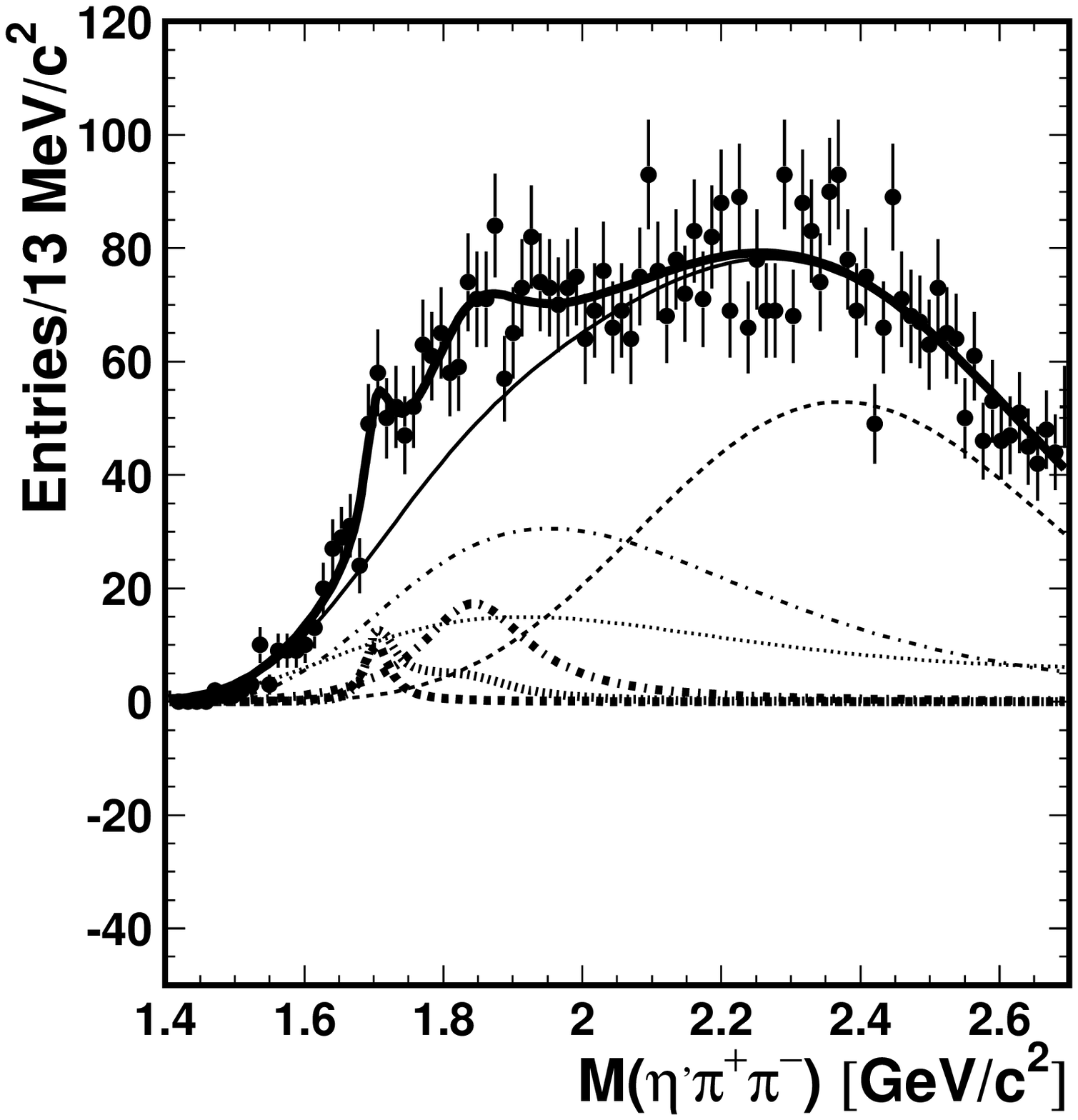,height=6.0cm,width=6.5cm}}}
\centerline{\hbox{\psfig{file=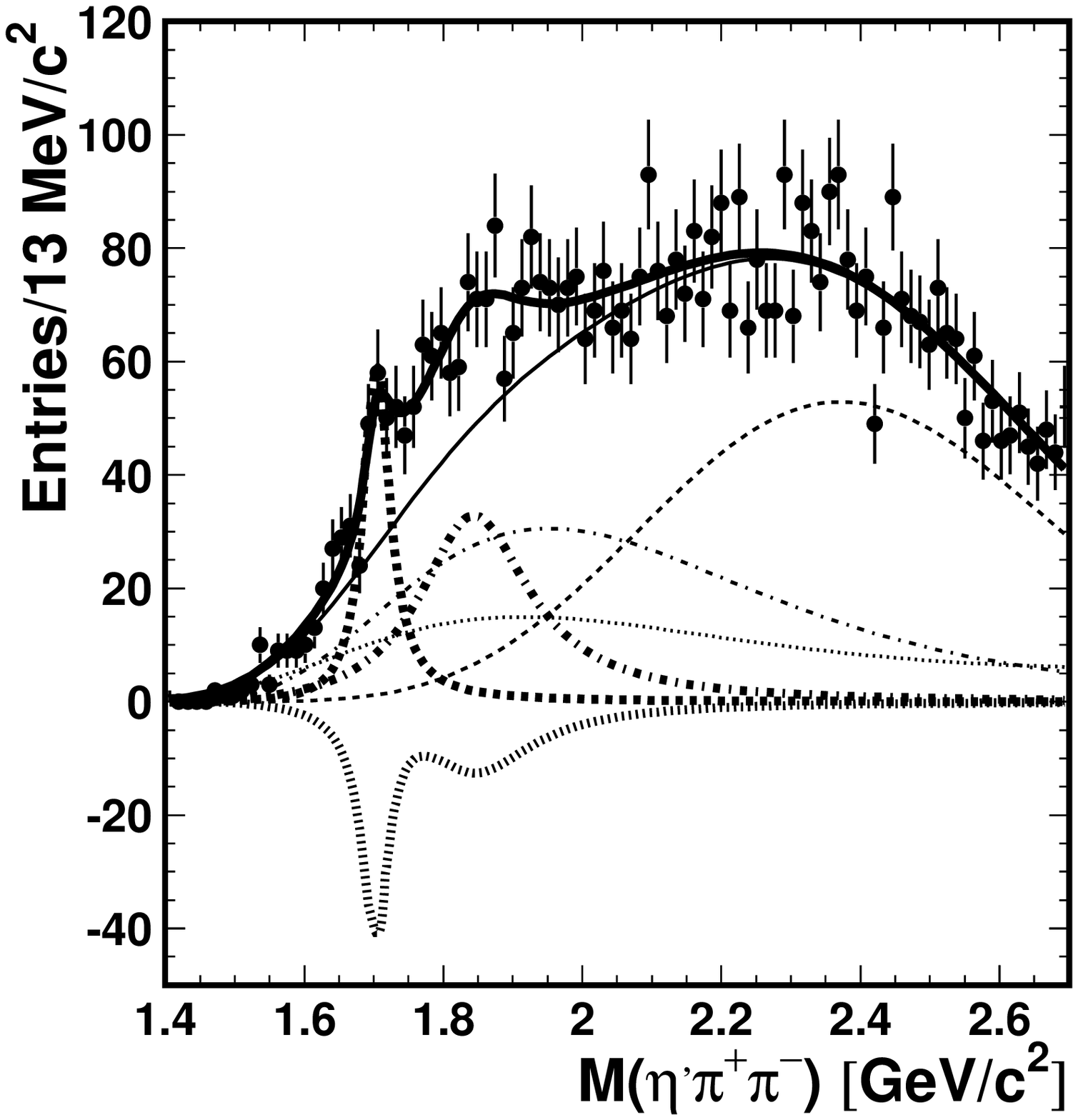,height=6.0cm,width=6.5cm}}}
\caption{Results of a combined fit for the $X(1835)$ and $\eta(1760)$
with interference between them.
The points with error bars are data. 
The thick solid line is the fit; the thin solid line is the total 
background. 
The thick dashed (dot-dashed, dotted) line is the fitted signal for the 
$\eta(1760)$ ($X(1835)$, the interference term between them). 
The thin dashed, dot-dashed and dotted lines are the $NR$, $b_1$ and
$b_2$ background components, respectively.
The upper (lower) panel represents the solution with
constructive (destructive) interference.
}\label{Fig fit result for x1835}
\end{figure}

Under the assumption of non-interference between the $R$ and $NR$ components, 
a fit with the $X(1835)$ and $\eta(1760)$ signals plus their interference
is performed to the lower-mass events.
Here, the $X(1835)$ mass and width are fixed at the BES values~\cite{BES X1835}. 
We find two solutions with equally good fit quality and 
the same $\eta(1760)$ mass and width; the results
are shown in Fig.~\ref{Fig fit result for x1835}. In either solution, 
the statistical significance is
$2.9\sigma$ for  the $X(1835)$ and $4.1\sigma$ for the $\eta(1760)$. 
The relative phase between the two resonances is determined to be 
$\phi_1 = (287^{+42}_{-51})^\circ$ for the constructive-interference solution and
$\phi_2 = (139^{+19}_{-9})^\circ$  for the destructive-interference one. 
The signal yields for the two solutions are determined to be
$Y_1 = 332^{+140}_{-122}$ and $Y_2 = 632^{+224}_{-231}$ for the $X(1835)$ 
and 
$Y_1 = 52^{+35}_{-20}$ and $Y_2 = 315^{+223}_{-165}$ 
for the $\eta(1760)$. 
The fitted mass and width of the $\eta(1760)$ 
are $M=(1703^{+12}_{-11})$ MeV$/c^2$ and $\Gamma =(42^{+36}_{-22})$ MeV$/c^2$.  
Upper limits on the product $\Gamma_{\GG} \BR(\etap\pip\pim)$ for the $X(1835)$
at the $90\%$ confidence level are determined to be 
$35.6$ eV$/c^2$ and $83$ eV$/c^2$
for the constructive- and destructive-interference solutions,
respectively.
The upper limit for the signal yield at $90\%$ confidence level is determined by
integrating the likelihood distribution 
convolved with a Gaussian function to include the systematic error.

Another fit without interference between the resonances is performed
to examine the significance of the $X(1835)$ signal.
The statistical significance from the fit with two incoherent resonances is 
found to be $3.2\sigma$ for the $X(1835)$ and $4.4\sigma$ 
for the $\eta(1760)$.
The $\eta(1760)$ mass and width are fitted to be 
$M=(1707.7^{+8.7}_{-7.0})$ MeV$/c^2$
and $\Gamma =(45^{+34}_{-21})$ MeV$/c^2$, respectively.
The products of the two-photon decay width and the branching fraction 
for the $X(1835)$ and $\eta(1760)$ decays to $\etap\pip\pim$ are estimated as
$\Gamma_{\GG}\mathcal{B}(X(1835)\ra\etap\pip\pim)~=~(23.1^{+6.3}_{-6.6})$ eV$/c^2$
and 
$\Gamma_{\GG}\mathcal{B}(\eta(1760)\ra\etap\pip\pim)~=~(6.7^{+2.8}_{-2.3})$ eV$/c^2$. 
The inclusion of the interference only mildly improves the fit.
The statistical significance of the interference term, defined as 
$\sqrt{-2\ln({\cal L}_{\rm no}/{\cal L}_{\rm yes})}$,
is $0.69\sigma$, where ${\cal L}_{\rm yes}$ (${\cal L}_{\rm no}$) is
the likelihood value of the fit with (without) interference.
There is a minor difference in the $\eta(1760)$ mass and width 
between the two fits with and without interference. 
The statistical significance of the $\eta(1760)$ mass difference between 
the fit result and the world-average value~\cite{pdg10}
is calculated as $\sqrt{-2\ln({\cal L}_{\rm fixed}/{\cal L}_{\rm free})}$,
and is found to be $2.6\sigma$ ($3.1\sigma$) for the two coherent (incoherent)
resonances. 
Here, ${\cal L}_{\rm fixed}$ and ${\cal L}_{\rm free}$
are the likelihood values of the fits with the $\eta(1760)$ mass fixed at the
world-average value and floating, respectively.

In the determination of the $X(1835)$ and $\eta(1760)$ significances, 
we have examined the effect of uncertainties of the following factors: 
(1) the $X(1835)$ mass or width varied by $\pm 1\sigma$; 
(2) a background fluctuation by changing the fit region;
(3) a background function that uses three threshold functions 
convolved 
with two-photon luminosity, efficiency and mass resolution;
(4) a fluctuation in the $b_1$ component by moving the $\etap$-$sdb$
selection mass window;
(5) a variation of $\pm 1\sigma$ in each of the background function parameters 
for the $b_1$ or $b_2$ components.
The fits of two incoherent resonances are performed under the variations
listed above.
The lowest (highest) significance $3.9\sigma$ ($5.0\sigma$) for the 
$\eta(1760)$ is obtained 
with the $X(1835)$ width increased (decreased) by $1\sigma$, 
while the significances under
the rest of variations are compatible with the values 
from the incoherent fit of $3.2\sigma$ 
for the $X(1835)$ and $4.4\sigma$ for the $\eta(1760)$. 
To ensure reliable estimation for the $X(1835)$, a fit with
floating masses and widths for both the $X(1835)$ and the $\eta(1760)$ is
performed. The yields, masses and widths are fitted to be
$Y=(444 \pm 158)$,
$M=(1833 \pm 30)$ MeV$/c^2$ and
$\Gamma=(202 \pm 66)$ MeV$/c^2$ for the $X(1835)$ and
$Y=(104 \pm 75)$,
$M=(1706.9 \pm 8.3)$ MeV$/c^2$ and
$\Gamma=(40 \pm 36)$ MeV$/c^2$ for the $\eta(1760)$.
In all variations, the fitted parameters for the $X(1835)$ are consistent with
those in the BES experiment.

\subsection{Angular distribution}\label{Angular distribution}

\begin{figure}[htbp]
\centerline{
\hbox{\psfig{file=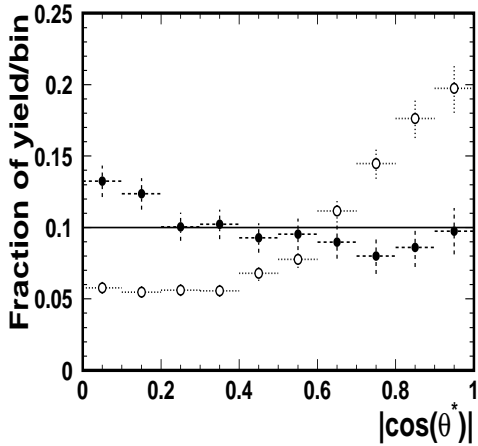,height=6.0cm,width=6.5cm}}
}
\caption{Angular distributions in the $\GG$ rest frame for two mass regions.
The solid circles are for the $X(1835)$ and $\eta(1760)$ region;
the open circles are for the $NR$ component in the upper sideband region. 
The yield in each bin is corrected for the 
efficiency and normalized to the sum of the corrected yield.
The horizontal line at $y = 0.1$ represents
an isotropic MC distribution.
} \label{cth in R1R2 region}
\end{figure}

\begin{figure}[htbp]
\centerline{
\hbox{\psfig{file=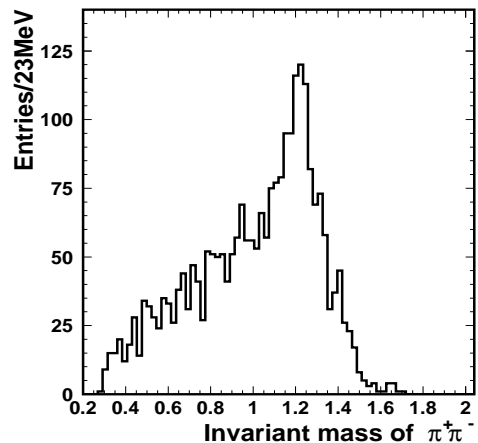,height=6.0cm,width=6.5cm}}
}
\caption{Invariant mass distribution of $\pip\pim$ 
for the $\etap\pip\pim$ events
selected in the upper sideband region of 
2.2 GeV$/c^2$ $<W<$ 2.7 GeV$/c^2$, where
a large signal for $f_2(1270)\ra\pip\pim$ decays is shown. 
} \label{2pi mass in NR region}
\end{figure}

We examined the distribution of $\theta^*$, the angle
between the $\etap$ momentum and the beam direction 
in the $\GG$ rest frame. 
The angular distribution is determined from $R$ and $NR$ yields extracted
from fitting the $\sumpt$ distribution 
sliced into ten angular bins
for the mass region of the $X(1835)$ and $\eta(1760)$ ($W < 2.04$ GeV$/c^2$) 
and its upper sideband ($W\in(2.2, 2.7)$ GeV$/c^2$). 
The distribution in the upper sideband region shows 
forward and backward peaks characteristic of a
higher-angular-momentum component,
which indicates strong contributions from the $\etap f_2(1270)$ production
(see Fig.~\ref{cth in R1R2 region}).
Indeed, a large $f_2(1270)$ signal is observed in the $\pip\pim$ invariant 
mass distribution for the $\etap\pip\pim$ events selected in that region,
as shown in Fig.~\ref{2pi mass in NR region}.
The dominant $\etap f_2(1270)$ component in the upper sideband region shows
interesting dynamics with a broad structure with favored quantum numbers 
$J^P = 2^+$.
A nearly isotropic distribution in the mass region below $2.04$ GeV$/c^2$ 
after the efficiency correction (with $\chi^2/N_{dof}=9.9/9$)
is compatible with the assumption of pseudoscalar quantum numbers 
for the $\eta(1760)$ and $X(1835)$.
However, a possible non-flat distribution for the $NR$ 
will influence
the distribution for the $R$ component; thus, a plausible 
$J^P$ value for each $R$ should be
examined with the $NR$ component subtracted once the existence of the $\eta(1760)$ and 
$X(1835)$ production is clarified.
No significant intermediate state is seen in the mass region 
below $2.04$ GeV$/c^2$. 
However, a minor contribution from another $J^P~=~2^-$ 
resonance~\cite{Crystal Ball} 
cannot be ruled out.

\subsection{Results of the $\etac$ fit}
\begin{figure}[htbp]
\centerline{
\hbox{\psfig{file=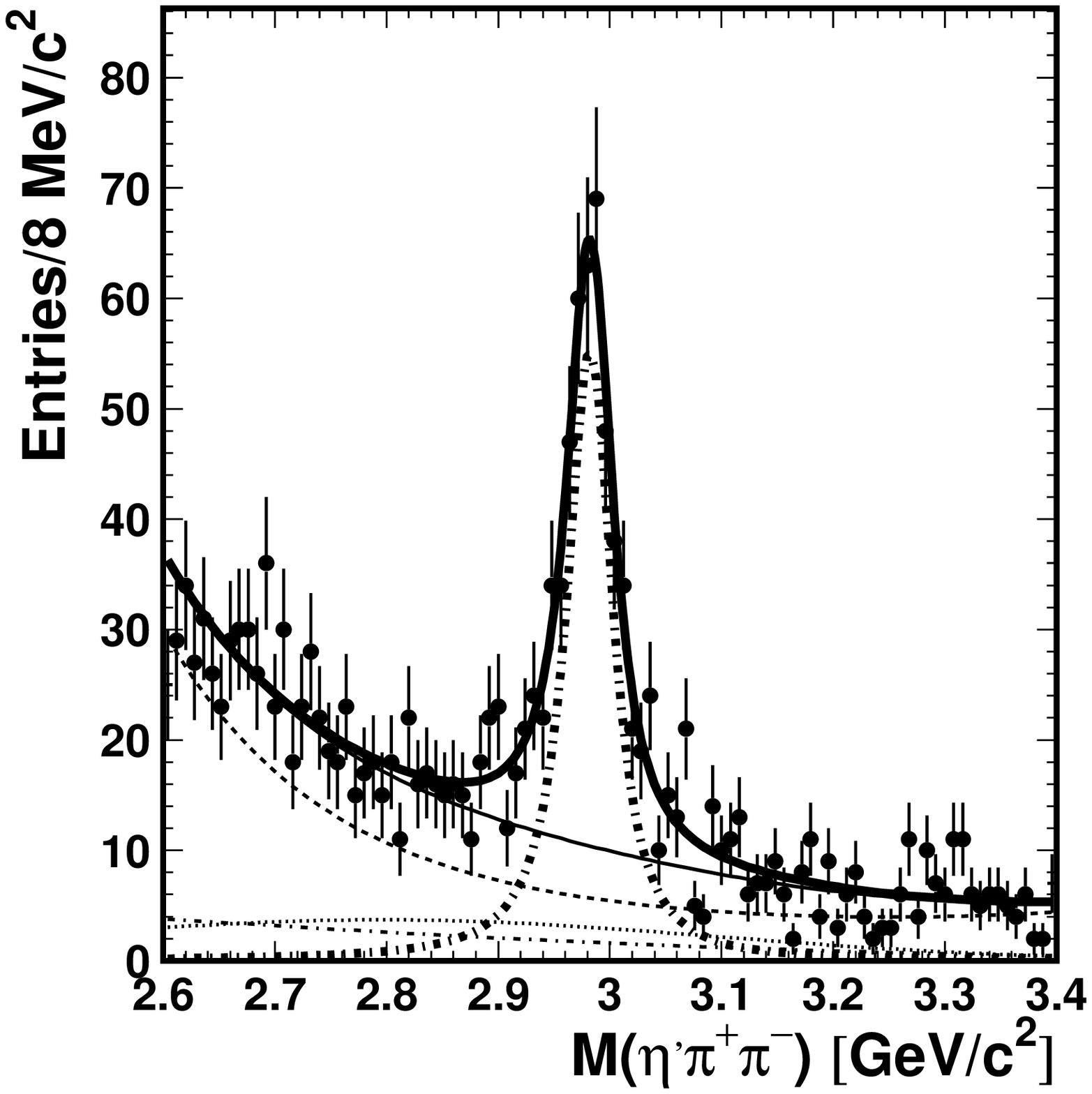,height=6.0cm,width=6.5cm}} }
\caption{The invariant mass distribution for the $\etap\pip\pim$ 
candidates in the mass range above $2.6$ GeV$/c^2$. The points with error 
bars are data. The thick-solid line is the fit; the thin-solid line is the 
total background. The thick dot-dashed line is the fitted signal for the 
$\etac$. The thin dashed, dot-dashed and dotted lines are the $NR$, $b_1$
and $b_2$ background components, respectively. 
}\label{Fig fit to etac}
\end{figure}

We first assume that there is no interference between the $\etac$ 
and the $NR$ background. 
Figure~\ref{Fig fit to etac} shows the $\eta^\prime \pip\pim$  invariant
mass distribution for the candidates with 
mass greater than 2.6 GeV/$c^2$ together with the fitted signal and 
background curves. 
The $\etac$ mass and width are determined to be
$M = (2982.7\pm1.8)$ MeV$/c^2$ and $\Gamma = (37.8^{+5.8}_{-5.3})$ MeV$/c^2$. 
The product of the
two-photon decay width and branching fraction for the $\etac$ is calculated 
using Eq.~(\ref{Eq1 prd of Br and 2g-width}).
Using the fitted $\etac$ signal yield of $n_{s} = 486^{+40}_{-39}$, we determine 
$\Gamma_{\GG} \BR(\etac\ra\etap\pip\pim) = (50.5^{+4.2}_{-4.1})$ eV$/c^2 $.

We now address the effect of possible  
interference between the $\etac$ resonance, hereafter referred to as $R$,
and the non-resonant component.
A precise description of the data in this case is impossible
without a good understanding of the background. 
As discussed in section~V-C,
the $NR$ component in the mass region above $2.2$ GeV$/c^2$
has a contamination of events from non-$0^-$ production via
two-photon processes. 
Although contamination is evident even in the $\etac$ mass region,
our data sample is insufficient to determine the type and 
rate of production of the non-$0^-$ states in this mass region.
The $NR$ component in our analysis can be subdivided into 
two types: one for the non-resonant final state (denoted as $NR1$) that 
interferes with $R$ and the other for production of 
various non-$0^-$ states (denoted $NR2$) that do not interfere 
with the $R$. 
The amplitude for $R$ production with interference with the $NR1$ term is written as
\begin{eqnarray}
\mathcal{M}(W) = A\cdot g_{BW}(W)\cdot e^{i\phi} + A_{NR1}\cdot g_{NR1}(W),
\label{BW intf to NR}
\end{eqnarray}
where $g_{BW}$ is the $BW$ function in Eq.~(\ref{BW function}),
$g_{NR1}$ is assumed to be a real function for $NR1$,
$\phi$ is the interference phase, and
$A$ and $A_{NR1}$ are normalization factors.
Assuming that $NR1$ and $NR2$ have the same shape,
the fitting function in Eq.~(\ref{Likelihood func})
for the $R$ and $NR$ components---where $R$ interferes with $NR1$ 
but not with $NR2$---can be expressed as
\begin{eqnarray}
f = n_s \cdot f_{s}(u; M,\Gamma) + 
n_{NR}\cdot f_{NR}(u; p_{NR}) +
f_{int},
\label{pdf NR and NR2 1}
\end{eqnarray}
where the interference term is
\begin{eqnarray}
f_{int} &=& 2\sqrt{\alpha_{NR}\cdot n_{NR}\cdot f_{NR}(u; p_{NR})} \nonumber \\
&& \cdot \sqrt{n_s \cdot f_{s}(u; M,\Gamma)} \cdot {\rm cos}(\theta+\phi)
\label{pdf NR and NR2 2}
\end{eqnarray}
with $\alpha_{NR} = n_1/n_{NR}$, $n_{NR} = n_1 + n_2$, and
$n_1$ and $n_2$ are the number of $NR1$ and $NR2$ events, respectively.
An intrinsic phase $\theta$ is determined by the $R$ mass, width and $W$
value.
The function $f$, including the $f_{int}$ term, is convolved with 
a mass resolution function after corrections for $dL_{\GG}/dW$ and efficiency.
The $f_s$ and $f_{NR}$ PDFs are normalized; the function $f_{int}$ is fully determined
by the fit parameters.

To investigate the possible effect of interference with the $NR$ component, 
a fit to the $\etac$ 
signal with interference with $NR1$ but without interference with $NR2$ 
is performed for various initial input values for the $\alpha_{NR}$ 
and $\phi$ parameters.
For the $\etac$, 
the fit gives two solutions with almost the same maximum likelihood value; the
mass and width of the $\etac$ are $M=2982.7$ ($2983.0$) MeV$/c^2$ 
and $\Gamma=36.4$ MeV$/c^2$ at $\alpha_{NR}=0.01\%$ ($100\%$);
these are quite consistent
with the result of the fit without interference. 
The differences in the $\etac$ mass and width with and without interference,
$\Delta M = 0.3$ MeV$/c^2$ and $\Delta \Gamma = 1.4$ MeV$/c^2$, are taken as 
model-dependent uncertainties in the determination of the mass and width.
However, the fits give very different values for the $\etac$ yield. 
If, for example,  $\alpha_{NR}$ is fixed at $100\%$ in the fit with interference,
the yields obtained are
$Y_1 = 854 \pm 59$ with $\phi_1 = (-92 \pm 5)^\circ$ for destructive interference
and 
$Y_2 = 264 \pm 22$ with $\phi_2 = ( 91 \pm 8)^\circ$ for constructive interference,
while the $\etac$ yield
of the incoherent fit is $486^{+40}_{-39}$. 
A strong correlation between $\alpha_{NR}$ and $\phi$ is observed from the fits:
$\phi_1$ and $\phi_2$ are close to 
$180^\circ$ and $-180^\circ$ ($90^\circ$ and $-90^\circ$), respectively,
if $\alpha_{NR}$ is close to zero ($100\%$). The insensitivity of
the maximum likelihood value for the fits in the full 
$\alpha_{NR}$ region
between zero and $100\%$ and a strong correlation between 
$\alpha_{NR}$ and $\phi$ 
imply large uncertainties in the determination of $\alpha_{NR}$, $\phi$ and the
strength of the interference term. 
With an additional error source from the interference term,
the $\etac$ yield has also a large uncertainty ranging from $264\pm 22$ 
to $486\pm 40$ for constructive interference and from $486\pm 40$ to $854\pm 59$ 
for destructive interference depending on the true $\alpha_{NR}$ and $\phi$ values.
Our fit results, as well as the absence
of any visual asymmetry in the $\etac$ line shape in the data, indicate that
the interference term cannot be determined without independent information 
on the $NR1$ component such as its angular distribution 
in the $\etac$ sideband mass region.   
The measured mass and width of the $\etac$ have a marginal dependence on
the interference, while the yield is strongly
correlated with 
the interference component and, thus,
cannot be determined precisely with the existing data sample.
The situation would improve if the interference effect were determined 
independently with a much larger data sample.

\section{Systematic errors}\label{Syst error}
To examine a possible bias in the mass measurement for 
the decay $R\ra\eta^\prime\pip\pim$, 
a data sample of $D^0\ra\etap K^0_{S}$ decays 
with $K^0_{S}\ra\pip\pim$ is selected 
with tight mass window requirements for the $\eta$ and $\etap$. The $D^0$ 
mass resulting from a fit of the invariant mass spectrum of 
$\etap K^0_S$ is lower than its nominal value
by 1.4 MeV$/c^2$, which is taken as an uncertainty of the mass scale
after a linear correction for mass value. 
The uncertainty in the width determination can arise from 
a difference in the mass resolution between data and MC simulation. 
This is estimated 
by changing the mass resolution by $\pm 1$ MeV$/c^2$ and is found to be
$2.0$ MeV$/c^2$ for the $\etac$ and $10$ MeV$/c^2$ for the $\eta(1760)$. 
Systematic errors on the mass, width and $\Gamma_{\GG}{\mathcal B}$
product 
due to uncertainties in the $NR$ background estimation 
are determined by varying 
the fit mass interval and $\sumpt$ requirement separately.
The error contributions from uncertainties in determination of the $b_1$ 
and $b_2$ backgrounds are minor for the $\etac$ but are sizable in the mass 
region below 2 GeV$/c^2$. 
The uncertainties in the resonance parameters, estimated by 
varying the shape parameters and yields of the $b_1$ and 
$b_2$ backgrounds by $\pm 1\sigma$ and added in quadrature, 
are taken as the corresponding errors 
for the $X(1835)$ and $\eta(1760)$, respectively.

There are additional sources of systematic errors 
in the $\Gamma_{\GG}{\mathcal B}$  product determination.
The trigger efficiency for four-track events is relatively high
because of redundant two-track and multi-track triggers in the Belle
first-level trigger. From the trigger simulation program, the difference 
in the efficiency with and without both trigger conditions satisfied is found  
to be $1\%$ ($2.7\%$) at an invariant mass of 2.98 (1.84) GeV$/c^2$; this is included
as a systematic error.
The efficiency for the pion identification, determined by using 
the inclusive $D^*$ sample, is lower than that from
MC simulation by $(1.40\pm 0.64)\%$ for the $\etac$ and  $(0.02\pm 0.60)\%$ 
for the $\eta(1760)$, and the corresponding contributions 
to the systematic error 
are $1.5\%$ and $0.6\%$, respectively.
The reconstruction efficiency for $\eta\ra\GG$ is studied with 
an inclusive $\eta$ sample, and its deviation from the MC simulation 
plus its error in quadrature is $4.9\%$. 
The uncertainty in the track reconstruction efficiency is $5.5\%$ and
that of the $\piz$-veto requirement is $3\%$. 
The accuracy of the two-photon luminosity function calculated by the TREPS
generator is estimated to be about $5\%$ including the error from neglecting
radiative corrections ($2\%$), the uncertainty from the form factor effect 
($2\%$), and the error of the total integrated luminosity 
($1.4\%$)~\cite{TREPS MC code}. 
The background contribution from the initial-state radiation processes is 
negligible~\cite{Charmonium two-photon}.
Furthermore, the run-dependent background conditions add an
additional uncertainty of $3\%$ in the yield determination.
A dominant source of systematic errors for the $X(1835)$ yield is the 
uncertainty of its decay width. It is estimated to be $18\%$ by changing
the width by $\pm 1\sigma_\Gamma$ in the fit for the yield extraction.

The systematic errors in the measurements of the mass and width 
for the $\etac$ and
$\eta(1760)$, as well as of the product $\Gamma_{\GG} \mathcal{B}$ 
for each resonance, 
are summarized in Table~\ref{Table sys err}. 

\begin{center}
\begin{table*}
\caption{Summary of systematic uncertainty contributions to the mass and width
for the $\etac$ and $\eta(1760)$ and to $\Gamma_{\GG} \mathcal{B}$ 
for the $\etac$, $\eta(1760)$ and $X(1835)$. 1-$R$ and 2-$R$ denote
one and two resonances in the fit, respectively.
}
\begin{tabular}{ccccc}
\hline\hline
Source          &~~~$\etac$~~~~~~&   \multicolumn{2}{c}{~~~$\eta(1760)$~~~}&~~~$X(1835)$~~~    \\\hline
                &\multicolumn{2}{c}{~~~1-$R$ fit}  &\multicolumn{2}{c}{~~~2-$R$ fit~~~}   \\\hline
\multicolumn{5}{c}{\emph{$\triangle(M)$ ({\rm MeV}$/c^2)$}}      \\\hline
Mass scale      &~~~2.2~~~~~~    &  \multicolumn{2}{c}{1.3}   &     -      \\
Background shape&~~~0.1~~~~~~    &   8       &  ~~~ 0.5       &     -      \\
$\etap$ sideband and $b_{any}$ 
                &~~~0.0~~~~~~    &  3.9      &  ~~~ 0.2       &     -      \\
$\sumpt$ requirement   
                &~~~0.4~~~~~~    &  4.5      &  ~~~ 0.6       &     -      \\
$X(1835)$ Width &     -~~~       &   -       &  ~~~ 0.9       &     -      \\
Total           &~~~2.2~~~~~~    &  10       &  ~~~ 1.8       &     -      \\\hline\hline
\multicolumn{5}{c}{\emph{$\triangle(\Gamma)$ ({\rm MeV}$/c^2)$}}       \\\hline
Mass resolution &~~~2.0~~~~~~    &  10       &  ~~~ 1.5       &     -      \\
Background shape&~~~1.9~~~~~~    &   7       &  ~~~ 6         &     -      \\
$\etap$ sideband and $b_{any}$ 
                &~~~0.02~~~~~~   &  17       &  ~~~ 7.1       &     -      \\
$\sumpt$ requirement   
                &~~~0.4~~~~~~    &  14       &  ~~~ 9         &     -      \\
$X(1835)$ Width &   -~~~         &    -      &  ~~~ 8         &     -      \\
Total           &~~~2.8~~~~~~    &   25      &  ~~~ 15        &     -      \\\hline\hline
\multicolumn{5}{c}{\emph{$\triangle(\Gamma_{\GG}\BR)/(\Gamma_{\GG}\BR)$ $(\%)$ }}\\\hline
$X(1835)$ Width       &  -~~~       &    -      & ~~~ 16      & 18         \\
Background shape      &~~~4.6~~~~~~ &    2      & ~~~ 13      & 2.6        \\
$\etap$ sideband and $b_{any}$      
                      &~~~0.03~~~~~~&   7.3     & ~~~ 15      & 3.8        \\
$\sumpt$ requirement   
                      &~~~2.2~~~~~~ &   0.6     & ~~~ 6.9     & 6.3        \\
Trigger efficiency    &~~~1~~~~~~   & \multicolumn{3}{c}{2.7~~~~~~~~~~}               \\
$\pi$ ID efficiency   &~~~1.5~~~~~~ & \multicolumn{3}{c}{0.6~~~~~~~~~~}               \\
$\eta$ rec. efficiency& \multicolumn{4}{c}{4.9}                                  \\
Track rec. efficiency & \multicolumn{4}{c}{5.5}                                  \\
$\piz$ veto           & \multicolumn{4}{c}{3}                                    \\
Two-photon Luminosity & \multicolumn{4}{c}{5}                                    \\
Run dependence        & \multicolumn{4}{c}{3}                                    \\
Total                 &~~~11~~~~~~  &   13      &   ~~~ 28    & 22                     \\\hline\hline
\end{tabular}
\label{Table sys err}
\end{table*}
\end{center}

\section{Results and discussion}
The results for the yields, masses 
and widths, as well as the product decay widths
are summarized in Table~\ref{Table etac result} for the $\etac$ 
and in Table~\ref{Table X1835 and eta1760 results} for the $\eta(1760)$ and 
$X(1835)$.

\begin{table}[htbp]
\caption{Summary of the results for the $\etac$:
$M$ and $\Gamma$ are the mass and width;
$Y$ is the yield;
$\mathcal{B}$ is the branching fraction for $\etac\ra\eta^\prime\pip\pim$;
$\Gamma_{\GG} \mathcal{B}$ is the product of the two-photon decay width 
and the branching fraction.
The world-average values are shown for comparison.
}
\begin{center}
\begin{tabular}{ccc}
\hline\hline
Parameters                          &    This work                        &  PDG             \\\hline
$Y$                                 &  $486^{+40}_{-39}\pm 53$            &                    \\
$M$,~MeV$/c^2$                      &  $2982.7\pm1.8\pm2.2$               & $2980.3\pm 1.2$    \\
$\Gamma$,~MeV$/c^2$                 &  $37.8^{+5.8}_{-5.3}\pm2.8 $        & $26.7\pm 3$        \\
$\Gamma_{\GG} \mathcal{B}$,~eV$/c^2$ 
                                    & $50.5^{+4.2}_{-4.1} \pm 5.6$        & $194\pm 97$ \\
$\mathcal{B},~\%$                   & $0.87\pm 0.20 $                     & $2.7 \pm 1.1$       \\\hline\hline
\end{tabular}
\end{center}
\label{Table etac result}
\end{table}

\begin{center}
\begin{table*}
\caption{Summary of the results for $\eta(1760)$ and $X(1835)$:
$M$ and $\Gamma$ are the mass and width;
$Y$ is the yield;
$\Gamma_{\GG} \mathcal{B}$ is the product of the two-photon decay width 
and branching fraction;
$Y_{90}$ and 
${(\Gamma_{\GG} \mathcal{B})}_{90}$ are the upper limits 
at $90\%$ confidence level with systematic error included. 
The $\eta(1760)$ mass and width from the two-resonance fit with 
interference, as well as
world average values, are shown for comparison. 
$S$ is the signal significance including systematic errors.
}
\begin{tabular}{ccccc}
\hline\hline
Parameter           & One resonance         &\multicolumn{2}{c}{Two interfering resonances}                     & Reference           \\\cline{3-4}
                    &                       & Solution I                             & Solution II              &                     \\\hline
                                         \multicolumn{5}{c}{\emph{$X(1835)$}}                                                         \\\hline
$M$,~MeV$/c^2$      &                       &  \multicolumn{2}{c}{1836.5 (fixed)}                               & $1836.5\pm3.0^{+5.6}_{-2.1}$~\cite{BES X1835}\\
$\Gamma$,~MeV$/c^2$ &                       &  \multicolumn{2}{c}{190 (fixed)}                                  & $190 \pm 9^{+38}_{-36}$~\cite{BES X1835}\\
$Y$                 &                       & $332^{+140}_{-122}\pm73$               & $632^{+224}_{-231}\pm139$&                    \\
$Y_{90}$            &                       & $<~650$                                & $<~1490$                 &                    \\
$\Gamma_{\GG} \mathcal{B}$,~eV$/c^2$          
                    &                       & $18.2^{+7.7}_{-6.7}\pm 4.0$            & $35^{+12}_{-13}\pm 8$ &                       \\
${(\Gamma_{\GG} \mathcal{B})}_{90}$~eV$/c^2$ 
                    &                       & $<~35.6$                               & $<~83$                   &                    \\
$S$,~$\sigma$       &                       &  \multicolumn{2}{c}{2.8}                                          &                    \\\hline\hline
                                  \multicolumn{5}{c}{\emph{$\eta(1760)$}}                                                            \\\hline
$M$,~MeV$/c^2$      &$1768^{+24}_{-25}\pm 10$&     \multicolumn{2}{c}{$1703^{+12}_{-11}\pm 1.8$}                & $1756\pm 9$~\cite{pdg10}        \\
$\Gamma$,~MeV$/c^2$ & $224^{+62}_{-56}\pm 25$&     \multicolumn{2}{c}{$42^{+36}_{-22}\pm 15$}                   & $96\pm 70$ ~\cite{pdg10}        \\
$Y$                 &$465^{+131}_{-124}\pm60$& $52^{+35}_{-20}\pm 15$                & $315^{+223}_{-165}\pm 88$&                    \\
$\Gamma_{\GG} \mathcal{B}$,~eV$/c^2$
                    & $28.2^{+7.9}_{-7.5}\pm3.7$& $3.0^{+2.0}_{-1.2}\pm 0.8$         & $18^{+13}_{-10}\pm 5$    &                    \\
$S$,~$\sigma$       &        4.7            &        \multicolumn{2}{c}{4.1}                                    &                    \\
$\phi$              &                       &$(287^{+42}_{-51})^\circ$                   &  $(139^{+19}_{-9})^\circ$    &                    \\\hline\hline
\end{tabular}
\label{Table X1835 and eta1760 results}
\end{table*}
\end{center}

The $\etac$ mass and width are measured to be 
$M=(2982.7\pm1.8(stat)\pm 2.2(syst) \pm 0.3(model))$ MeV$/c^2$ and 
$\Gamma = (37.8^{+5.8}_{-5.3}(stat)\pm 2.8(syst) \pm 1.4(model))$ MeV$/c^2$, and
are consistent with the recent results from BES~\cite{BES etac} and
Belle~\cite{vinokur}. 
If we assume that there is no interference, the directly measured 
product for the $\etac$ decay width to $\etap\pip\pim$
is determined to be
$\Gamma_{\GG} \mathcal{B}(\etac\ra\etap\pip\pim)=(50.5^{+4.2}_{-4.1}\pm 5.6)$ eV$/c^2$, 
which is marginally consistent with the existing value $(194\pm 97)$ eV$/c^2$ from 
the indirect measurements~\cite{pdg10}. 
Instead of a direct reference to the world-average value for 
$\Gamma_{\GG}(\etac)$, 
we determine it from the ratio of 
$\Gamma_{\GG}\Gamma(K\overline{K}\pi)/\Gamma_{\rm total}$ 
$= (0.407\pm 0.027)$ keV$/c^2$
to
$\Gamma(K\overline{K}\pi)/\Gamma_{\rm total}$
$= (7.0\pm 1.2)\times 10^{-2}$~\cite{pdg10}, and obtain
the width $\Gamma_{\GG}(\etac)$ $= (5.8\pm 1.1)$ keV$/c^2$
with a smaller relative error. With that as an input,
the branching fraction is calculated to be
$\mathcal{B}(\etac\ra\etap\pip\pim) = (0.87\pm 0.20)\%$, where
both statistical and systematic errors are included.

We report the first evidence for $\eta(1760)$ decay to
$\etap\pip\pim$ and
find two solutions for its parameters, depending on the
inclusion or not of the $X(1835)$, whose existence is marginal in our fits. 
The decay $\eta(1760)\ra\etap\pip\pim$ is found
with a significance of $4.7\sigma$ including systematic error, 
with the assumption that the $X(1835)$ is not produced;
the $\eta(1760)$ mass and width are determined to be
$M=(1768^{+24}_{-25}\pm 10)$ MeV$/c^2$ and 
$\Gamma = (224^{+62}_{-56}\pm 25)$ MeV$/c^2$.
The fitted $\eta(1760)$ mass is consistent with the existing measurements
~\cite{DM2 eta1760, BES eta1760}.
The product of the two-photon decay width and the branching fraction 
for the $\eta(1760)$ decay to $\etap\pip\pim$ is determined to be
$\Gamma_{\GG} \mathcal{B}(\eta(1760)\ra\etap\pip\pim) = 
(28.2^{+7.9}_{-7.5}\pm 3.7)$ eV$/c^2$.
When the mass spectrum is fitted with two coherent resonances, 
the $\eta(1760)$ and $X(1835)$,
the $\eta(1760)$ mass and width are found to be 
$M = (1703^{+12}_{-11}\pm 1.8)$ MeV$/c^2$ and 
$\Gamma=(42^{+36}_{-22}\pm 15)$ MeV $/c^2$, and
the signal significances including the systematic error estimated to be 
$4.1\sigma$ for the $\eta(1760)$ and
$2.8\sigma$ for the $X(1835)$. 
Upper limits on the product $\Gamma_{\GG} \BR$ for the $X(1835)$ decay
to $\etap\pip\pim$ at the $90\%$ confidence level for two fit solutions 
are determined:
$\Gamma_{\GG} \BR(X(1835)\ra\etap\pip\pim)~<~35.6$ eV$/c^2$ 
with $\phi_1=(287^{+42}_{-51})^\circ$ for constructive interference
and $\Gamma_{\GG} \BR(X(1835)\ra\etap\pip\pim)~<~83$ eV$/c^2$ 
with $\phi_2=(139^{+19}_{-9})^\circ$ for destructive interference.

In summary, we report the first observation of $\eta^{\prime}\pi^+\pi^-$
production in two-photon collisions. We measure the mass, width and the 
product of the two-photon width and the branching fraction for the 
$\eta_c$. 
We also report the first evidence for the $\eta^{\prime}\pi^+\pi^-$ decay 
mode of the $\eta(1760)$.
No strong evidence for the $X(1835)$ is found.

\begin{acknowledgments}
We extend our special thanks to J.X.~Wang of IHEP (Beijing)
for many helpful discussions.
We thank the KEKB group for the excellent operation of the
accelerator; the KEK cryogenics group for the efficient
operation of the solenoid; and the KEK computer group,
the National Institute of Informatics, and the 
PNNL/EMSL computing group for valuable computing
and SINET4 network support.  We acknowledge support from
the Ministry of Education, Culture, Sports, Science, and
Technology (MEXT) of Japan, the Japan Society for the 
Promotion of Science (JSPS), and the Tau-Lepton Physics 
Research Center of Nagoya University; 
the Australian Research Council and the Australian 
Department of Industry, Innovation, Science and Research;
the National Natural Science Foundation of China under
contract No.~10575109, 10775142, 10875115 and 10825524; 
the Ministry of Education, Youth and Sports of the Czech 
Republic under contract No.~LA10033 and MSM0021620859;
the Department of Science and Technology of India; 
the Istituto Nazionale di Fisica Nucleare of Italy; 
the BK21 and WCU program of the Ministry Education Science and
Technology, National Research Foundation of Korea,
and GSDC of the Korea Institute of Science and Technology Information;
the Polish Ministry of Science and Higher Education;
the Ministry of Education and Science of the Russian
Federation and the Russian Federal Agency for Atomic Energy;
the Slovenian Research Agency;  the Swiss
National Science Foundation; the National Science Council
and the Ministry of Education of Taiwan; and the U.S.\
Department of Energy and the National Science Foundation.
This work is supported by a Grant-in-Aid from MEXT for 
Science Research in a Priority Area (``New Development of 
Flavor Physics''), and from JSPS for Creative Scientific 
Research (``Evolution of Tau-lepton Physics'').
\end{acknowledgments}


\begin{thebibliography}{dd}
\bibitem{pdg10} K. Nakamura {\em et al.} (Particle Data Group), 
J. Phys. G {\bf 37}, 075021 (2010).

\bibitem{Charmonium two-photon} S.~Uehara {\em et al.} (Belle Collaboration), 
Eur. Phys. J. C {\bf 53}, 1 (2008).

\bibitem{BES etac} M.~Ablikim {\em et al.} (BES Collaboration), 
Phys. Rev. Lett. {\bf 108}, 222002 (2012).

\bibitem{bram} N.~Brambilla {\em et al.}, Eur. Phys. J. C {\bf 71},
1534 (2011).

\bibitem{vinokur} A.~Vinokurova {\em et al.}
(Belle Collaboration),  Phys. Lett. B {\bf 706}, 139 (2011).

\bibitem{BES X1835} M.~Ablikim {\em et al.} (BES Collaboration), 
Phys. Rev. Lett. {\bf 106}, 072002 (2011);
Phys. Rev. Lett. {\bf 95}, 262001 (2005).

\bibitem{S.L. Shi and C.S. Gao} S.L.~Zhu and C.S.~Gao, 
Commun. Theor. Phys. {\bf 46}, 291 (2006); 
Z.G.~Wang and S.L.~Wan, J. Phys. G {\bf 34}, 505 (2007). 

\bibitem{G.J Ding and M.L. Yan} G.J.~Ding and M.L.~Yan, 
Eur. Phys. J. A {\bf 28}, 351 (2006).

\bibitem{N. Kochelev and D. Min} N.~Kochelev and D.P.~Min, 
Phys. Rev. D {\bf 72}, 097502 (2005);
Phys. Lett. B {\bf 633}, 283 (2006).
\bibitem{Bin An Li} B.A. Li, Phys. Rev. D {\bf 74}, 034019 (2006).
\bibitem{X.G. He} X.G.~He, X.Q.~Li, X.~Li and J.P.~Ma, 
Eur. Phys. J. C {\bf 49}, 731 (2007).

\bibitem{Tao Huang} T.~Huang and S.L.~Zhu, Phys. Rev. D {\bf 73}, 014023 (2006).
\bibitem{E Klempt} E.~Klempt and A.~Zaitsev, Phys. Rept. {\bf 454}, 1 (2007). 

\bibitem{BES ppbar enhancement} J.Z.~Bai {\em et al.} (BES Collaboration), Phys.
Rev. Lett. {\bf 91}, 022001 (2003).

\bibitem{BESIII x1835 ppbar}  M.~Ablikim {\em et al.} (BES Collaboration),
Chin. Phys. C {\bf 34}, 421 (2010).

\bibitem{MK3 eta1760} R.M.~Baltrusaitis {\em et al.} (MARKIII Collaboration),
 Phys. Rev. Lett. {\bf 55}, 1723 (1985); Phys. Rev. D {\bf 33}, 1222 (1986).

\bibitem{DM2 eta1760} D.~Bisello {\em et al.} (DM2 Collaboration),
 Phys. Rev. D {\bf 39}, 701 (1989); Phys. Lett. B {\bf 192}, 239 (1987).

\bibitem{BES eta1760} M.~Ablikim {\em et al.} (BES Collaboration),
 Phys. Rev. D {\bf 73}, 112007 (2006); 
J.Z.~Bai {\em et al.} (BES Collaboration), Phys. Lett. B {\bf 446}, 356 (1999).

\bibitem{Page and Li} P.R.~Page and X.Q.~Li, 
Eur. Phys. J. C {\bf 1}, 579 (1998).

\bibitem{J.L. Rosner} J.L.~Rosner, AIP Conf. Proc. {\bf 815}, 218 (2006).

\bibitem{Belle detector} A.~Abashian {\em et al.} (Belle Collaboration),
Nucl. Instrum. Methods Phys. Res., Sect. A {\bf 479}, 117 (2002).

\bibitem{KEKB collider} S.~Kurokawa and E.~Kikutani, Nucl. Instrum.
Methods Phys. Res. Sect. A {\bf 499}, 1 (2003), and other papers included in
this volume.

\bibitem{TREPS MC code} S.~Uehara, KEK Report 96-11 (1996).

\bibitem{Lgg BW for 2gam} V.M.~Budnev, I.F.~Ginzburg, 
G.V.~Meledin and V.G.~Serbo,
Phys. Rep. C {\bf 15}, 181 (1975);
J.~Field, Nucl. Phys. B {\bf 168}, 477 (1980), and Erratum-ibid, B {\bf 176}, 545 (1980). 

\bibitem{JETSET} T.~Sj\"ostrand, Comput. Phys. Commun. {\bf 82}, 74 (1994).

\bibitem{Belle MC simulation} The detector response is simulated 
with GEANT, R.~Brun 
{\it et al.}, GEANT 3.21, CERN Report DD/EE/84-1, 1984.

\bibitem{backg any} S.~Uehara {\em et al.} (Belle Collaboration), 
Phys. Rev. D {\bf 82}, 114031 (2010).

\bibitem{etac and etappipi} H.~Nakazawa (Belle Collaboration), 
Nucl. Phys. Proc. Suppl. {\bf 124}, 220 (2008).

\bibitem{mnfit} I.C.~Brock, A Fitting and Plotting Package Using
MINUIT, version 4.07, Dec. 22th, 2000. 
The threshold function in MINUIT is defined as
$f_{thresh}(x) = A\cdot (x-x_0)^p e^{c_1(x-x_0)+c_2(x-x_0)^2}$,
where A, p, $c_1$, $c_2$ and $x_0$ are parameters.

\bibitem{Crystal Ball} K.~Karch {\em et al.} (Crystal Ball Collaboration),
 Z. Phys. C {\bf 54}, 33 (1992).

\end{thebibliography}
\end{document}